\documentclass{raa}
\usepackage{graphicx,times}
\usepackage{natbib}
\usepackage{amssymb,amsmath}
 \usepackage[T1]{fontenc}
  \usepackage{aecompl}
\bibpunct{(}{)}{;}{a}{}{,}
\usepackage[a4paper=true,dvipdfm=true,pagebackref=true]{hyperref}
\hypersetup{pdftitle = The title of my PDF, pdfauthor = My name, pdfsubject= The subject, pdfkeywords = keyword1 keyword2 keyword3}
\hypersetup{colorlinks = true, linkcolor = green, anchorcolor = red, citecolor = blue, filecolor = red, pagecolor = red, urlcolor = red}

\begin{document}

   \title{Astronomical Data Fusion Tool Based on PostgreSQL
$^*$ \footnotetext{\small $*$ Supported by the National Natural
Science Foundation of China.} }

 \volnopage{ {\bf 2016} Vol.\ {\bf X} No. {\bf XX}, 000--000}
   \setcounter{page}{1}

   \author{Bo Han\inst{1}, Yanxia Zhang\inst{2}, Shoubo Zhong\inst{1,2},Yongheng Zhao\inst{2}}
   \institute{International School of Software, Wuhan University, Wuhan, 430072, P.R.China; \\
   \and  Key Laboratory of Optical Astronomy, National Astronomical Observatories,
Chinese Academy of Sciences, 20A Datun Road, Chaoyang District,
100012, Beijing, P.R.China  {\it zyx@bao.ac.cn}\\
\vs \no
   {\small Received 2015 June 12; accepted }
}

\abstract{With the application of advanced astronomical technologies,
equipments and methods all over the world, astronomy covers from
radio, infrared, visible light, ultraviolet, X-ray and gamma ray
band, and enters into the era of full wavelength astronomy. How to
effectively integrate data from different ground- and space-based
observation equipments, different observers, different bands,
different observation time, requires the data fusion technology. In
this paper we introduce the cross-match tool that is developed by
the Python language and based on the PostgreSQL database and uses
Q3C as the core index, facilitating the cross-match work of massive
astronomical data. It provides four different cross-match functions,
namely: I) cross-match of custom error range; II) cross-match of
catalog error; III) cross-match based on the elliptic error range;
IV) cross-match of the nearest algorithm. The cross-match result set
provides good foundation for subsequent data mining and statistics
based on multiwavelength data. The most advantage of this tool is a
user-oriented tool applied locally by users. By means of this tool,
users may easily create their own databases, manage their own data
and cross-match databases according to their requirements.  In addition, this tool is also able
to transfer data from one database into another database. More
importantly, the tool is easy to get started and used by astronomers
without writing any code.
}

   \authorrunning{B. Han et al. }            %author_head in even pages
   \titlerunning{Astronomical Fusion Tool}  % title_head in odd pages
   \maketitle

\section{Introduction}
In astronomy, the data fusion technology is the foundation of
multiwavelength astronomical research. Through the catalog fusion,
we can integrate multiple independent catalogs or images, databases
or data sets by positions or object names and other information into
a whole, so as to deepen the understanding of the celestial bodies
and promote the discovery of new objects or new physical phenomena.
For example, Metchev et~al. (2008) reported new L and T dwarfs in a
cross-match of SDSS and 2MASS. Maselli et~al. (2015) found new blazars by cross-matching
the recent multi-frequent catalogs. Multiwavelength data obtained are of great significance for further
statistical analysis and data mining (e.g. Zhang \& Zhao, 2003, 2004; Gao et~al. 2009; Zhang et~al. 2013). However, for the catalog
fusion technology, the most important step is the calculation of
cross-match, which refers to finding the corresponding entry in a
catalog to each source in another catalog by means of the source
location as the center. In general, the observation has errors due to
various factors, which causes difficulty for cross-match. Any source
in different catalogs has one or more counterparts within a
certain error radius.

In recent years, there are many studies focusing on the development
of cross-match tools. Thus cross-match tools are in bloom.

VizieR (http://vizier.u-strasbg.fr/), operated at CDS, Strasbourg,
France, provides access to the most complete library of published
astronomical catalogs and data tables available on line organized
in a self-documented database. Query tools allow the user to select
relevant data tables and to extract and format records matching
given criteria. But the cross-match work only supports a small
number of records.

SIMBAD (http://simbad.u-strasbg.fr/simbad/) is an astronomical
database operated at CDS, Strasbourg, France, which provides basic
data, cross-identifications, bibliography and measurements for
astronomical objects outside the solar system (Wenger et~al. 2000).
SIMBAD has many kinds of query modes, such as object name,
coordinates and various criteria. SIMBAD also provides links to some
other on-line services. Users may submit lists of objects and
scripts to query. Similar to VizieR, the number of lists cannot be
large.

The NASA Extragalactic Database (NED,
http://www.ned.ipac.caltech.edu/), managed by NASA, contains names,
positions, and a variety of other data for extragalactic objects, as
well as bibliographic references to published papers, and notes from
catalogs and other publications. NED may be searched for objects in
many ways, including by name, position, redshift, type, or by object
classifications. NED also offers a number of other tools and
services. If users want to query a large number of objects, users
may submit a NED Batch Job, and retrieve the results at Pick Up
Batch Job Results. NED provides another batch query, i.e. one RA and
Dec position or object name per line, maximum 500 positions and/or
object names per request.

The Tool for OPerations on Catalogues And Tables (TOPCAT,
http://www.star.bris.ac.\\uk/$\sim$mbt/topcat/) is an interactive
graphical viewer and editor for tabular data (Taylor 2005). It
offers a variety of ways to view and analyze tables, including a
browser for the cell data themselves, viewers for information about
table and column metadata, and facilities for sophisticated
interactive 1-, 2-, 3- and higher-dimensional visualization,
calculating statistics and joining tables using flexible matching
algorithms. It is developed in Java language and limited by computer
memory when running. When cross-matching very large tables or tables
with lots of columns, the computer is inclined to be out of memory.
TOPCAT's sister package is STILTS (The Starlink Tables
Infrastructure Library Tool Set), which is based on STIL, the
Starlink Tables Infrastructure Library. STILTS offers many of the
same functions as TOPCAT and forms the command-line counterpart of
the GUI table analysis tool TOPCAT. STILTS is robust, fully
documented, and designed for efficiency, especially with very large
datasets.

The CDS cross-match service (http://cdsxmatch.u-strasbg.fr/xmatch/)
is a new data fusion and data management tool, which is used to
efficiently cross-identify sources between very large catalogs
(all VizeR tables, SIMBAD) or between a user-uploaded list of
positions and a large catalog (Boch et al. 2012). About the xMatch algorithm, please
refers to Pineau et~al. (2011). Users interact with the CDS xMatch
service through a Web application. Due to narrow network bandwidth,
it has some limitations, for example, long jobs are aborted if
computation exceeds 100 minutes while short jobs are aborted if
computation exceeds 15 minutes; the search radius is maximized to
120 arcsec for a simple cross-match; the cone radius is no more than
15 degrees for a cone search; results are saved no more than 7 days
following their submission. Moreover the total size of uploaded
tables is limited to 100 MB for anonymous users, and 500 MB for
registered users.

2MASS catalog server kit, developed by Yamauchi (2011), realizes a
high-performance database server for the 2MASS Point Source Catalog
and several all-sky catalogs. This kit uses the open-source
PostgreSQL, adopts an orthogonal $xyz$ coordinate system as the
database index and applies other techniques (table partitioning,
composite expression index, and optimization in stored functions) to
enable high-speed search and cross-match of huge catalogs.

Pei (2011) and Pei et al. (2011) had developed a high-efficient large-scale
catalogue oriented fusion toolset based on MySQL database and HTM index. Zhang et al.
(2012) had developed the toolkit for automated database creation and cross-match task,
with which users may create their own databases and easily cross-match catalogs.
Although the cross-match speed is quick, it costs a long time to
retrieve the cross-matched result. In other words, the second
operation of matched result is necessary before application.

TAPVizieR is a new way to access the VizieR database using the ADQL
language and the TAP protocol (Landais et~al. 2013). The database is
based on PostgreSQL and the sky indexation depends on HEALPix.
TAPVizieR provides query and cross-match functions. The result
access is only limited to the owner recognized by the IP address and
saved no more than 5 days. The execution time of an ADQL query is
limited to 5 hours.

The Large Survey Database (LSD,
http://research.majuric.org/trac/wiki/LargeSurvey\\Database) is a
framework for storing, cross-matching, querying, and rapidly
iterating through large survey datasets (catalogs of $>10^9$ rows,
$>1$ TB) on multi-core machines. It is implemented in Python,
written and maintained by Mario Juric. LSD applies nested butterfly
HEALPix pixelization and the catalogs are partitioned into cells in
space and time. LSD employs LSD/MapReduce as the high-performance
programming model.

To some extent, all cross identification tools strongly promote the
study of multiwavelength astronomy and provide strong cross-match
function. Nevertheless these tools are almost for large data
managers; some are for astronomers; some are difficult to learn;
some are accessed by network and limited by network bandwidth, so
users can't upload too large catalogs. Considering all factors, we
develop a user-oriented cross-match tool based on PostgreSQL
database with the sky-indexing Q3C, and continuously improve the
system efficiency of identification, making the amount of
cross-match records increase to tens of millions of lines and even
higher. At the same time, we also develop the auxiliary tool of
data exchange between different databases, helping astronomers better manage data.

\section{Methods and implementation}

We develop a cross-match tool based on PostgreSQL database and Q3C
index. This tool has a friendly user interface for convenient use.
Moreover we design an auxiliary tool applied for data exchange
between different databases (MySQL, PostgreSQL). The detailed
development scheme of the cross-match tool is described in the
following.

\subsection{Database architecture}

The PostgreSQL (http://www.postgresql.org/) database has rich
functions and a very stable version, and its default
security configuration gets quite a lot of security personnel's
praise. The consistency of the SQL specification and data integrity
only allow interacting with the database through rigorous way, which
has the very good guarantee for high quality of data. PostgreSQL is
``object/relation'' database management system, and it has the
characteristics of open-source, multi-platform support and free use
etc. From the academic opinions, it has very advanced and creative
characteristics, and very rich third party libraries. It
focuses on academic study. It also has very good support in
astronomy, for example, the third party libraries PGSphere and Q3C
developed for astronomy play a very good promotion for the study of
astronomy. We choose Q3C index in the PostgreSQL database, and then
use the psycopg2 interface program based on the Python language to
interact with the PostgreSQL database and operate the data.

\subsection{Q3C}

Quad Tree Cube (Q3C, http://code.google.com/p/q3c/) is the plugin
for PostgreSQL database and developed for dealing with large
astronomical catalogs or any catalog of objects on the sphere. Q3C
provides some functions, for example, fast circular, elliptical or
polygonal searches on the sphere, fast positional cross-match and
nearest neighbor queries (Koposov \& Bartunov 2006). The base of
this scheme is the cube inscribed in the sphere, each face of the
cube is a quad tree structure. The quad tree structure establishes
the mapping of 2D coordinates in the square to the bitmask. There
are six faces, so the 3 bits pointing to the face number are
generated. Therefore the mapping of the cube to the integer numbers
are created. Each point of sphere has a corresponding integer value,
which is called as IPIX value. There are many adjacent IPIX values
around the center, which can realize quick search for the surface of
sphere by creating indexes with the IPIX values. In order to
effectively make use of each index, each spatial query will be split
in the pixel. And each pixel represents a continuous range of IPIX values.
It should be easy and rapid retrieval from the database
for part of the data of the sphere.

\subsection{Support of system environment}

These tools are coded by Python language. It is recommended to use
the Linux operating system. Users need to install the Python (the
general distribution of Linux will bring the Python program with
itself), and need the following database system (remote or local)
and third party libraries as support. 1) PostgreSQL (9.0+): support
for the latest SQL grammar, higher functional integrity. 2) MYSQL
(5.1+): the performance is very efficient. 3) Q3C: the plugin for
PostgreSQL database. The good virtual terminal is also embedded in
these tools, so the user can correct the run error of the program, according to the feedback of the virtual
terminal, making the program operate perfectly.

\subsection{The scheme of cross-match tool}

The cross-match tool based on Q3C partition scheme requires two
catalog tables (same or different) saved on the database server (local or
remote). But the database must be PostgreSQL which deploys the Q3C
index; otherwise the program will not work normally. Users should
ensure that the database supports remote access when choosing the
remote server. The operation process is shown in Figure 1.

\begin{figure}[h]
\center
\includegraphics[bb=0 0 567 229,width=10cm,clip]{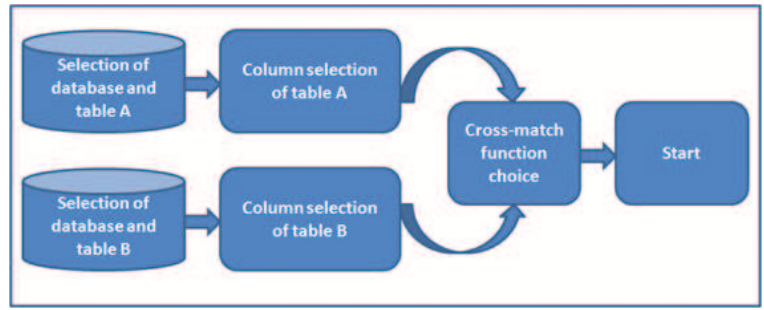}
  \caption{The scheme of cross-match tool}
\end{figure}

\subsection{Graphical user interface}

For convenient application of the cross-match tool, we develop a
Graphical User Interface (GUI) indicated in Figure 2. Users can
directly click the program code to run GUI with mouse, or can also manually
use the command in the command line to run GUI. GUI is separated
from the main program, which assists users to adjust the parameters
according to a certain standard that is used for the main program.
That is to say, the main program do not depend on GUI, the user can
manually select the related file to run the main program. As shown
in Figure~2, this cross-match tool provides four different
cross-match functions: I) cross-match with custom error range; II)
cross-match with catalog error; III) cross-match based on the
elliptic error range; IV) cross-match by the nearest algorithm. The
matched result may output only the matched records or all results.
When default setting of parameter list is adopted, the matched
result will give all parameters of matched sources from table A and
table B as well as distance between the matched sources. In addition,
users may choose the parameters they need. If all the required data
exist in the database, users can apply the tools to do the
cross-match work. After the cross-match work is finished, users can
find a matched result folder in the local directory, then the result
set exists in the folder, which saves all matched results or the
conflict doesn't happen. According to the start time of the
cross-match work, users may choose what they want.

\begin{figure}[h]
\center
\includegraphics[bb=0 0 555 628,width=10cm,clip]{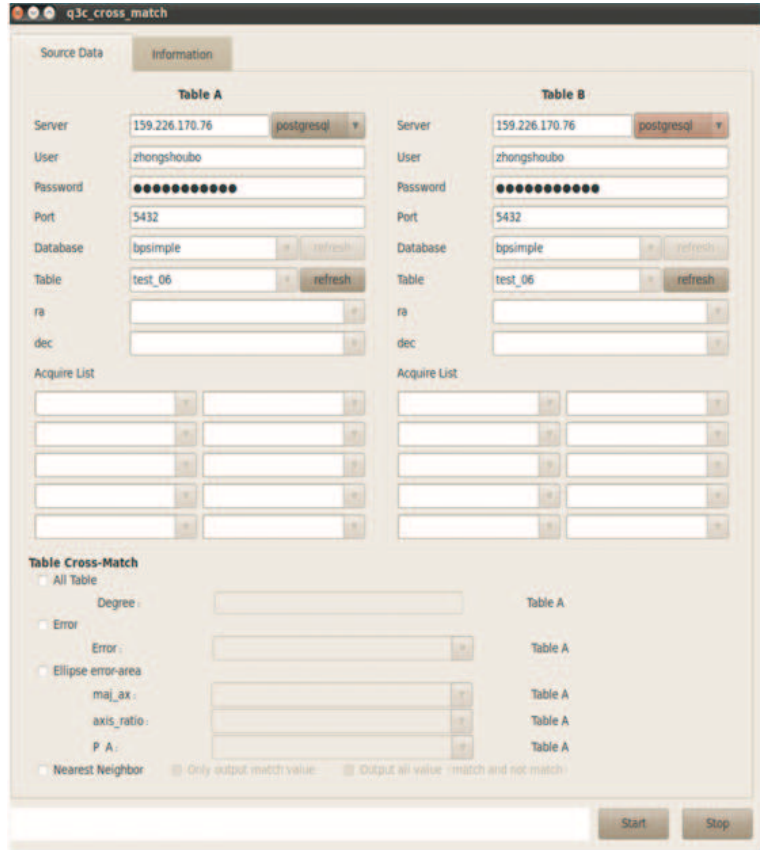}
  \caption{The operation interface of cross-match tool}
\end{figure}

In the process of developing the cross-match tool, we also develop an
auxiliary tool, which realizes the data exchange between different
databases MySQL and PostgreSQL, and automatically create the
required index, so that the astronomers can directly exchange data
between different databases and do not need to do extra work to
migrate data. It is convenient for astronomers to manage
the data. The user interface is shown in Figure~3.

\begin{figure}[h]
\center
\includegraphics[bb=0 0 516 363,width=10cm,clip]{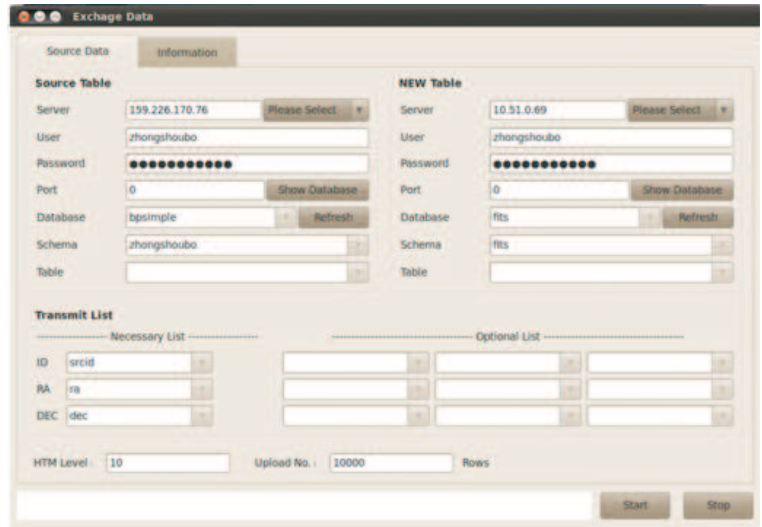}
  \caption{The operation interface of data exchange program}
\end{figure}

\section{The experimental results}

In order to confirm the reliability of the tool, we make a
simulation experiment. We randomly select the part of the Sloan Digital Sky Survey (SDSS) database to
do the experiment, a total of 10,000,000 rows as the source
catalog, and we respectively set the matching radius as 2.5 arcsec,
3 arcsec and 5 arcsec to perform the catalog cross-match with
itself. Our test platform is using two computers: one is a local
database and program running platform, and another is the platform
for the remote database that is accessed through fast ethernet. The
specific configurations are shown in Table~1.

\begin{table}[h]
\tiny
\begin{center}
\caption{The configuration of experiment platform software and
hardware}
\begin{tabular}{rllllll }
\hline \hline
      Platform&  CPU& Memory&  OS&  Core&  HDD(rpm)  &Software\\
\hline Database& Inter(R)Xeon(E5-4607)& 8G &Red Hat Enterprise Linux
& Linux version & 10,000&Mysql5.6
Postgresql9.3\\
& six-core at 2.20GHz&&& 2.6.9-22.ELsmp&&\\

 Program&    Inter(R)Xeon(E5630)  &8G &Ubuntu 10.04 & Linux version&
7,200&Python 2.6.2
MySQL5.6\\
&quad-cores at 2.53GHz&&& 2.6.28-15-generic&& PostgreSQL8.4\\
 \hline
\end{tabular}
\bigskip
\end{center}
\end{table}

The identification result is shown in Table 2. Table 2 indicates
that all data can match successfully, and no matter whether the matching
radius is 2.5 arcsec, 3 arcsec or 5 arcsec, the identification rate
is 100\% and the number of wrong matched records is 0, so we can
draw a conclusion that the cross-match program is fairly reliable.

\begin{table}[h]
\begin{center}
\caption{The results of simulation experiment}
\begin{tabular}{rlllll }
\hline \hline

Matching radius&   No. of identification results &Matched
rate& Wrong matched No.\\
(arcsec)&&\\
 \hline
2.5& 10,000,000 &  100.0\% &    0\\
3.0& 10,000,000 &  100.0\% &    0\\
5.0& 10,000,000 &  100.0\% &    0\\
\hline
 \hline
\end{tabular}
\bigskip
\end{center}
\end{table}

In order to prove effectiveness of our tool, we compare our tool
with the tool of Pei (2011) and Pei et~al. (2011). Under the
hardware configuration is fixed, the performance of cross-match is
mainly related to the size of catalog A and not to that of catalog
B. We choose different amounts of catalog A to do experiments.
Selecting 100,000, 4,000,000 and 10,000,000 records respectively from
SDSS database as catalog A and 10,000,000 records from the Wide-field Infrared Survey Explorer (WISE) database
as catalog B, the time spent on cross-match is listed in Table~3.
The time of Pei et al.'s tool used for cross-match consists of two
parts: one part for cross-match and another for getting parameters
from both catalog A and catalog B. Taking cross-match between
10,000,000 rows of SDSS and 10,000,000 rows of WISE for example, the
time cost on cross-match is 665 seconds and the time spent on
getting parameters is 2,980 seconds. The total time is 3,645
seconds. With the same data, our tool performs cross-match and
getting parameters at the same time and the spent time is 1,040
seconds. Obviously, our tool is much faster than that of Pei
et~al. (2011). The CPU occupation of Pei et al. (2011) and ours is 70\% and
92\%, respectively.

\begin{table}[h]
\begin{center}
\caption{Time spent on cross-match}
\begin{tabular}{rllllll }
\hline \hline
Method & No. of catalog A  &  No. of catalog B &   Time(s) &CPU\\
\hline our tool & 100,000  &   10,000,000 &   4   & 88.0\%\\
                &4,000,000 &   10,000,000 &   420 & 90.0\%\\
                &10,000,000&   10,000,000 &   1040& 92.0\%\\
\hline
Tool of Pei et al.&  100,000&   10,000,000 &    185&   72\%\\
                  &4,000,000&   10,000,000 &    435&   73\%\\
                  &10,000,000&  10,000,000 &    665&   70\%\\

\hline
 \hline
\end{tabular}
\bigskip
\end{center}
\end{table}

We also compare our tool with the more popular cross-match tools, such as CDS x-Match service and TOPCAT.
By means of the CDS x-Match service, the UKIRT Infrared Deep Sky Survey (UKIDSS) DR9 (82,655,526 rows) database is cross-matched with itself,
the running time spends 24 minutes. A small table uploaded by users is very quickly cross-matched with
those large tables provided by the CDS x-Match service in a few minutes and it is very fast
to download the result. Although the cross-match speed of the CDS x-Match
service is the fastest, the time to download cross-matched result is very long especially for large sky survey database due to internet bandwidth. Users only cross-match their
own tables with the tables provided by this service. Moreover the total size of uploaded tables is limited
to 100 MB for anonymous users, and 500 MB for registered users. In addition, if there are some cross-match tasks on the service, users
should wait in queue. With TOPCAT in our computer, we select the UKIDSS DR9 to cross-match with itself, the time reading data costs about 46 minutes and the time on cross-match
spends about 36 minutes. The advantages of TOPCAT are easy use and direct getting the result without second processing,
while the amount of tables to cross-match is limited by the memory of the user's computer due to pure JAVA program for TOPCAT.
Only in terms of cross-match speed, our tool is slower than the CDS x-Match service and TOPCAT. As far as ease of use and convenience,
the CDS x-Match service and TOPCAT are worthy of reference and learning. Compared to all these tools, our tool is a user-oriented
cross-match tool, by which users may manage and cross-match the
catalogs they need, moreover transfer data from one database (MySQL
or PostgreSQL) to another database (PostgreSQL or MySQL). By means
of the Graphical User Interface (GUI), users may be very easy to
apply this tool to perform cross-match. The merits of our tool are that users may automatedly create their own databases and cross-match the tables according to their requirements,
the tables may be stored in their own computers or servers, the cross-matched result can be also easily input to the database for ease management and later use.

In terms of hardware, the performance of cross-match is affected by
many factors: CPU, network, hard disk drive (HDD).
Therefore network bandwidth and a hard drive spin directly influence
the speed of cross-match with the same computation platform, and then the upgrading of network and HDD
accelerate cross-match. From the above experiments, the CPU
occupation is very high. When the data are large enough, especially
the amount of data adds up to 10,000,000, the CPU occupation is
92\%. If the number of cross-match data further increases, our tool
runs unstably based on our present personal computer. As a result,
we put forward a solution that the huge table should be split into a
number of smaller tables when the rows of a table is larger than
10,000,000. One advantage of PostgreSQL is to permit users a
seamless access to the split tables. The creation of child tables
adopts CHECK constraints of PostgreSQL. For much larger tables with
more than 10,000,000 records, our tool automatically splits them
into child tables according to source positions, and then users may
choose child tables to cross-match one by one.

At present, the bottleneck of any cross-match task is still on the I/O limit not on computation.
With the number of survey data increases, the I/O problem apparently comes out.
The improvement of I/O speed is the key of cross-match. Pineau et al. (2015) put forward some techniques to
strengthen cross-match performance, for instance, large tasks are split into smaller independent
pieces for scalability, matching speed accelerates due to multi-threading, sequential reads and several tree data-structures.
LSST project adopt a new kind of database, the Large Survey Database (LSD), which is a framework for storing, cross-matching, querying,
and rapidly iterating through large survey datasets on multi-core machines and it is implemented in Python. The new techniques
developed in the computer field may be borrowed into astronomy. Therefore the key problem faced in cross-match task will be solved.
In general, the cross-match refers to the spacial cross-match only considering the position information, in fact, there are many factors to
judge which one is the true corresponding one. Budavari \& Szalay (2008) presented a Bayesian approach for object matching by means
of not only spatial information, but also physical properties, such as colors, redshift, and luminosity. They provided a practical
recipe to implement an efficient recursive algorithm to evaluate the Bayes factor over a set of catalogs with known circular errors in positions.
Rots et al. (2009) also applied a Bayesian method to cross identify the Chandra
Source Catalog (CSC) with the SDSS DR7 source catalog, which takes into account the positions, position
errors, as well as the detailed footprints of the catalogs, and provided a Bayesian Factor and an associated probability for each match.
In terms of cross-match accuracy, there is still much work worth studying. In the present version, our tool hasn't performed the parallel
computing and distributed storage, and only implements spacial cross-match. In the future, our tool may improve from
all these respects, and thus the speed and accuracy of cross-match task both improve with the help of new developing approaches.

\section{Conclusion}

This paper is devoted to the development of efficient and
easy-to-use catalog cross-match tool. According to the
astronomical data characteristics of large-scale, multiband and
distributiveness, we analyze the international existing cross-match
tools and develop a simple and efficient data fusion tool according
to the actual needs. It could be used for the tens of millions of
lines of data, or even larger, and we adopt the method of process
pool to fully use CPU, which ensures the efficiency of cross-match.
Users can arbitrarily choose the cross-match functions to obtain the
final multiband cross-match result according to their own needs.
Given high identification speed and good accuracy, the tool is a
booster to large scale catalog cross-match. Moreover it is easy to
learn and use for astronomers. It has been of very important
significance for the subsequent data mining and statistical analysis
work. Nevertheless, there are still lots of functions needed to be
improved in the future. For example, this tool is limited to system
environment, only in the use of the environment based on PostgreSQL
database system. For the matched result set, users may choose less
than 10 columns of data in each table except the right ascension RA
and declination Dec, or all columns are output when setting default. If
users want to get more than 12 columns, they need adopt the default
setting. In the following work, the cross-match services can be
fully deployed on Web servers, letting users input data into our
database server, using our platform to do the cross-match work,
which will greatly reduce the workload of astronomical work.
Meanwhile the user can give feedback to us through the network
service in the use process, so our program will be more robust. In
the program itself, we can further improve the high efficiency of
program using the multiprocessing module to realize the parallel
program. The benefit is that it does not produce potential errors
but high efficiency, the direct operation of process usually will be
more efficient than the use of a good package library. So we can
consider using the Python subprocess module to initiate multiple
processes, but in this case each process will get their own unique
results set, then we still need secondary processing to get
the final result set. Database performance improvements can directly
improve the running efficiency of the program, when a tabular data
quantity is large enough, the memory reaches a bottleneck, even if
using the index, the query work can still be very slow. Here it is
very necessary for us to do the table division operation, dividing
the large table into a plurality of small tables to do the
cross-match work, so that the efficiency will be improved a lot.
Moreover, if the database is arranged by distributed cluster layout
method, the program robustness will be much higher. So in the future,
we will take many better ways (e.g. MapReduce, Spark) to improve the efficiency of
the program and apply new data mining methods to perform not only
spacial cross-match, but also physical matching. Altogether
the development of new computer technologies, new types of databases
and database index methods oriented to big data is of great importance for today's multiwavelength
observations and the time domain science of the upcoming survey telescopes.

\begin{center} \bf Acknowledgment\end{center}

We are very grateful to the referee for his insightful comments. This paper is funded by National Key Basic Research
Program of China 2014CB845700, National Natural Science Foundation
of China under grants NO.61272272, No.11178021, No.11033001,
NSFC-Texas A\&M University Joint Research Program No.11411120219. We
acknowledgment SDSS, WISE and UKIDSS databases.

\end{document}